\pdfoutput=1
\documentclass[preprint,12pt,3p]{elsarticle}

\usepackage{graphicx}
\usepackage{amsmath}
\usepackage{amssymb}
\usepackage{blindtext}
\usepackage{multirow}
\usepackage{color}
\usepackage{fancyhdr}
\usepackage{floatflt}
\usepackage{color,soul}
\usepackage{graphics}
\usepackage{subfigure}
\usepackage[ruled]{algorithm2e}
\usepackage{times}
\usepackage{hyperref}
\usepackage[colorinlistoftodos]{todonotes}

\journal{Electric Power Systems Research}

\begin{document}

\begin{frontmatter}

\title{A Backend Framework for the Efficient Management of Power System Measurements}

\author[label1]{Ben McCamish}
\author[label1]{Rich Meier}
\author[label2]{Jordan Landford}
\author[label2]{Robert B.~Bass}
\author[label3]{David Chiu}
\author[label1]{Eduardo Cotilla-Sanchez\corref{cor1}}
\ead{ecs@oregonstate.edu}
\cortext[cor1]{Corresponding author. Tel.:~+1(541)737-8926}

\address[label1]{School of Electrical Engineering \& Computer Science, Oregon State University, Corvallis, OR, USA}
\address[label2]{Maseeh College of Engineering \& Computer Science, Portland State University, Portland, OR, USA}
\address[label3]{Department of Mathematics \& Computer Science, University of Puget Sound, Tacoma, WA, USA}

\begin{abstract}
Increased adoption and deployment of phasor measurement units (PMU) has provided valuable fine-grained data over the grid. Analysis over these data can provide insight into the health of the grid, thereby improving control over operations. Realizing this data-driven control, however, requires validating, processing and storing massive amounts of PMU data. This paper describes a PMU data management system that supports input from multiple PMU data streams, features an event-detection algorithm, and provides an efficient method for retrieving archival data. The event-detection algorithm rapidly correlates multiple PMU data streams, providing details on events occurring within the power system. The event-detection algorithm feeds into a visualization component, allowing operators to recognize events as they occur. The indexing and data retrieval mechanism facilitates fast access to archived PMU data. Using this method, we achieved over $30\times$ speedup for queries with high selectivity. With the development of these two components, we have developed a system that allows efficient analysis of multiple time-aligned PMU data streams. 
\end{abstract}

\begin{keyword}
PMU \sep data management \sep bitmap index \sep electrical distance \sep correlation \sep power system contingency
\end{keyword}

\end{frontmatter}

% \linenumbers

\begin{abstract}
Increased adoption and deployment of phasor measurement units (PMU) has provided valuable fine-grained data over the grid. Analysis over these data can provide insight into the health of the grid, thereby improving control over operations. Realizing this data-driven control, however, requires validating, processing and storing massive amounts of PMU data. This paper describes a PMU data management system that supports input from multiple PMU data streams, features an event-detection algorithm, and provides an efficient method for retrieving archival data. The event-detection algorithm rapidly correlates multiple PMU data streams, providing details on events occurring within the power system. The event-detection algorithm feeds into a visualization component, allowing operators to recognize events as they occur. The indexing and data retrieval mechanism facilitates fast access to archived PMU data. Using this method, we achieved over $30\times$ speedup for queries with high selectivity. With the development of these two components, we have developed a system that allows efficient analysis of multiple time-aligned PMU data streams.
\end{abstract}

\section{Introduction}
\label{sec:intro}
Recently, power grid operations have been complicated by increased penetration of variable generation, load congestion, demand for quality electric power, environmental concerns, and threats to cyber-security and physical infrastructure. Pressure from these issues compel engineers to create tools that leverage modern communications, signal processing, and analytics to provide operators with insight into the operational state of power systems. As Horowitz, \textit{et al.} explained, there are multiple aspects to achieving the level of knowledge and control necessary to keep one of the world's greatest engineering feats stable and operational~\cite{Paper030}. To this end, utilities have been deploying phasor measurement units (PMU)\footnote{Also known as \textit{synchrophasors}, we refer to them as PMUs throughout this paper.} across the grid. At a high-level, PMUs are sensors that measure electrical waveforms at short fixed intervals~\cite{phadke-93-pmu}. A unique feature of PMUs is that they are equipped with global positioning systems (GPS), allowing multiple PMUs distributed in space to be synchronized across time. With a proper set of analytics put in place, the mass deployment of PMUs can offer utility operators a holistic and real-time sense of grid status.

With the recent deployment of PMUs on a large scale, their applications are growing. PMUs provide visibility over the grid at increasing speeds allowing for real-time monitoring of grid conditions~\cite{lin2014algorithm,gollav2014lav,salehi2012laboratory}. PMU placement is also being optimized to provide accurate information about the grid while minimizing the number of units required to achieve observability~\cite{li2013information}. Furthermore, this space has seen a significant increase in algorithms that aid in control and mitigation of grid operational issues. For example, efforts have emphasized using PMU data to monitor critical power paths~\cite{Paper036}, identify transmission line fault locations~\cite{Paper039}, isolate and mitigate low-frequency zonal oscillations~\cite{Paper040}, and predict critical slowing down of the network~\cite{Paper044}.

Despite increase in PMU use, there is still a lack of verification of the data generated by PMUs. Many algorithms assume input data streams to be robust, reliable, and available at all times. However, this is not the case in a real PMU network. Not only do corrupt data streams cause false positives during normal operation, but they reduce confidence in data generated during transient events. The standard for PMU measurements (IEEE C37.118.1-2011) provides some testing and error measurement specifications for these types of situations, but clarification of how a PMU should act is not stated~\cite{Misc005}. Some recent works, namely \cite{Paper045, Paper046, Ghanavati:2014}, have made some initial steps in verifying the output of PMU devices before informing the operation of higher-level power system control algorithms. They have specifically stressed the importance of data integrity during transient situations. These efforts, however, have not sufficiently solved the event-detection problem. 

A second issue not addressed in many of the above works is a result of the sophisticated nature of sensing and data gathering in today's PMUs. In the field, each PMU data stream is collected and coalesced by a device known as a phasor data concentrator (PDC) before being written to large, but slow, non-volatile storage, e.g., hard disks or tape. When data streams from many PMUs are combined, it can amount to massive volumes of data each year (on the order of 100s of TBs). Unfortunately, common data processing tasks, such as real-time event detection, \textit{ad hoc} querying, data retrieval for analysis, and visualization require scanning or randomly accessing large amounts of PMU data on disk. These tasks can require prohibitive amounts of time. Therefore, in addition to the identification problem stated above, there is also a significant data management problem that has thus far gone unaddressed. 

In this paper, we describe a framework for addressing both the inconsistent data (data-flagging) problem, as well as the back-end mechanisms that manage the massive PMU data streams. Our goal is to improve near-real-time event/error detection, data management, and archived data access in a manner that can inform higher level control operations and visualization for operator decision-making. To this end we have developed a system architecture capable of interchanging components. This paper presents our execution of these system components, providing the outcomes expected of this system.

The remainder of this paper is organized as follows. The following work will first depict the system architecture as a whole and how each component works in Section~\ref{sec:background}. Next, Section~\ref{sec:methodology} will describe the details of implementation of these components. The results from our experiments will be discussed in Section~\ref{sec:results}. Possible future expansion routes are highlighted in Section~\ref{sec:future}. Finally, we conclude this work in Section~\ref{sec:conclusion}

% \section{A real-time framework for PMU-assisted algorithms}
\section{System Design}
\label{sec:background}
% With the two problems laid out in Section~\ref{sec:intro}, our goal is to create a system that addresses these two issues. To this end, 
We have created a system that is composed of two primary components, \textit{Monitoring and Live Analysis} and \textit{Historical Data Management}. Within these components we developed two methods to fulfil these functions, a correlation matrix with a graphical display and a data management algorithm known as a \textit{bitmap index}. This system allows for sufficient validation of the PMU data while providing fast operator query support on the large database. 

\begin{figure}[ht]
\centering
f\includegraphics[width=.85\textwidth]{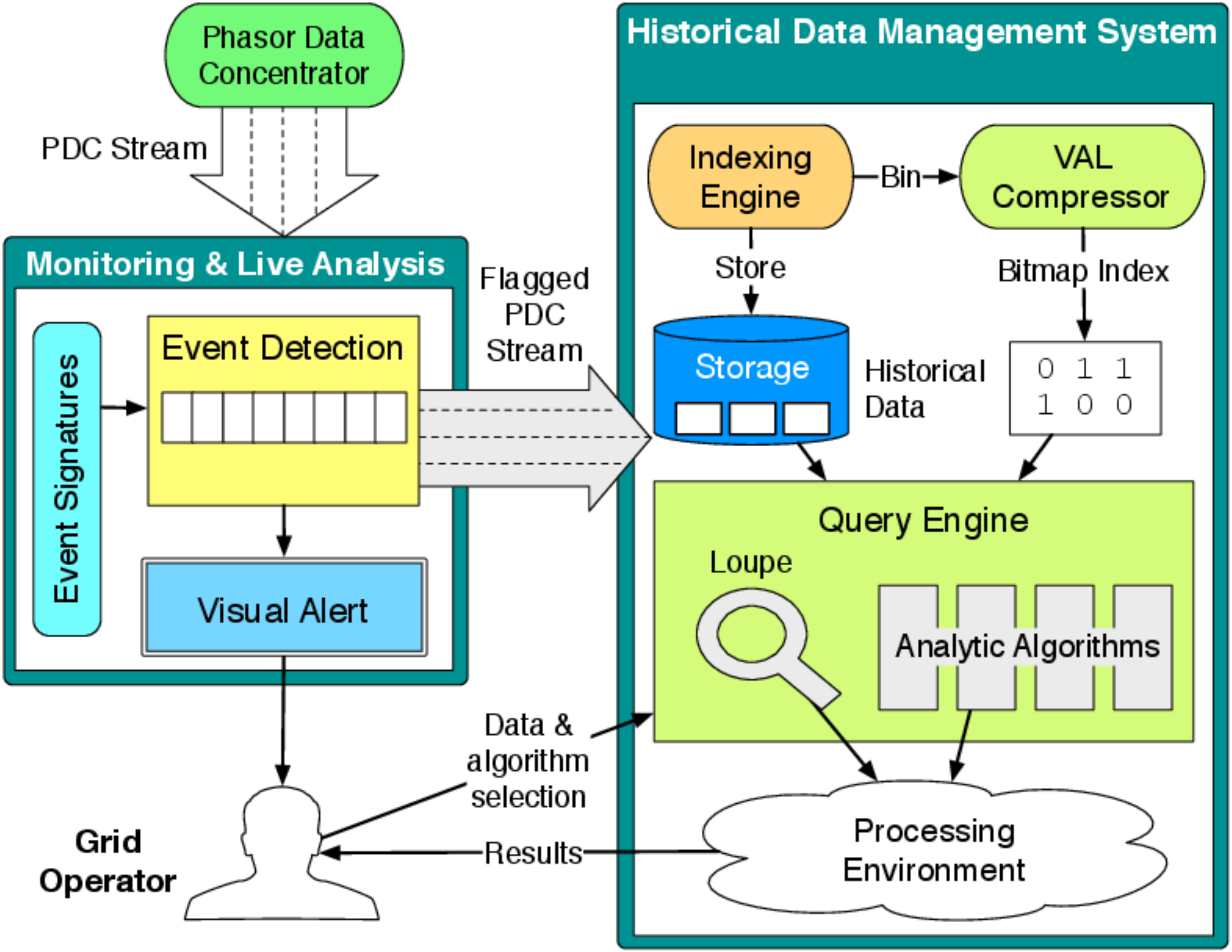}
\caption{System Architecture}
\label{fig:system}
\end{figure}

Figure~\ref{fig:system} illustrates the system architecture. Data arriving from a phasor data concentrator (PDC) is first given to the \textit{Monitoring and Live Analysis Subsystem}. This subsystem comprises three main components. The Event Detection engine inputs a set of known power-systems event signatures and analyzes the PDC stream in a single pass. To perform this one-pass analysis, we use a correlation matrix, which also provides visual alerts to the operator by depicting various event signatures. Using this correlation matrix we are able to detect and identify events occurring within the power grid monitored by the PMUs.

%\todo[inline]{dc: the high-level idea of correlating PMUs is lost in this section and in the intro. We talk about this correlation matrix, but we don't say what it's for. The details come later, but to a reader, they're kept in the dark until then. Someone more knowledgeable and eloquent than me should add this brief narrative.}

The PDC data is sent to the Historical Data Management System for archiving. First, data is discretized (binned) to generate a bitmap index (described in detail in the next subsection). The bitmap, once compressed, allows for efficient response to queries from the operator. This system architecture allows for the operator to monitor the grid in real time, including the ability to detect various power system events and data errors. While monitoring the grid the operator can query the large database of past PMU values using the Data Management subsystem, allowing for replay of historical events through the Monitoring and Live Analysis system or simply for further examination.

We believe that the Monitoring and Live Analysis subsystem, coupled with the Historical Data Management subsystem, may improve operator decision making. Being able to monitor the grid and detect events while having the capability to query past synchrophasor measurements grants the operator this capability.
%\todo[inline]{dc: there's a pretty sizeable imbalance between the background on the bitmap stuff on the live/monitoring stuff. I later found all that detail in Methodology section. I propose that we pull the bitmap stuff out of this section, and move it to the Methodology section. Instead, this section should serve only as a high-level ``tour'' of all the system components.}

\subsection{Historical Data Management System}
The Historical Data Management System uses the Bitmap Index method. Within this component, a different data management method can be utilized as well. Bitmap indices~\cite{ONeil:1987:bitmap}, are popular for managing large-scale data sets~\cite{Sinha07:multibitmap,RomosanSWMM13,su-hpdc2013,Fusco10:netfli,Wu06:DataWarehouse,hive}. A bitmap $B$ is an $m \times n$ matrix  where the $n$ columns represent range-bins, and the rows correspond to the $m$ tuples/records (e.g., PMU measurements).
% To represent data as a bitmap, each attribute is first partitioned into bins that denote either a value or a range of values.
A bit $b_{i,j} = 1$, if the $i$th record falls into the specified value/range of the $j$th bin, and $b_{i,j} = 0$, otherwise.

\begin{table}[ht]
\begin{center}
  \begin{tabular}{|l||l|l|l|l||l|l|l|}                          \hline
  \textbf{Records} & \multicolumn{7}{c|}{\textbf{Bins}} \\    \hline
    &\multicolumn{4}{c||}{$X$}&\multicolumn{3}{c|}{$Y$}\\
    &$x_1$&$x_2$&...&$x_{50}$&$y_1$&$y_2$&$y_3$\\                         \hline
    $t_1$&0&1&...&0&0&0&1\\
    $t_2$&0&0&...&0&0&1&0\\
    $t_3$&0&0&...&1&0&0&1\\                                         
    $...$&$...$&$...$&$...$&$...$&$...$&$...$&$...$\\                                         \hline
    \end{tabular}
\end{center}
\caption{\label{tbl:bitmap}An Example Bitmap Index}
\end{table}

Consider the bitmap in Table~\ref{tbl:bitmap}. Suppose these example data have two attributes, $X$ and $Y$, the values of $X$ are known to be integers in the range $(0,50]$, and that the values of $Y$ can be any real number. Due to its small cardinality, we can generate a bin $x_j$ for each possible value of $X$. The values of $Y$ are, however, continuous and unbounded. We therefore discretize its values, i.e., decide on an appropriate cardinality of bins to represent $Y$ and select the range of values associated with each bin. In our example, we chose to use only three bins, $y_1 = (-\infty, -5]$, $y_2 = (-5, 5)$, and $y_3 = [5,\infty)$. 

Suppose we want to retrieve all records from disk where $X < 25$ and $Y = 0$. We can identify the candidate records by computing the following boolean expression,
\begin{gather*}
v_R = (x_1 \vee ... \vee x_{24}) \wedge y_2 
\end{gather*}
% $v_R$ is a bit-vector, where the value of $i$th bit denotes whether the $i$th record in the file is promising. Therefore, 
The bits with a value of $1$ in $v_R$ correspond with the set of candidate records on disk,
\begin{gather*}
R = \{t~|~(t[X] < 25) \wedge (-5 < t[Y] < 5)\}
\end{gather*}
Intuitively, there could be false positives in $R$, which requires checking, but only the records $r_i \in R$ with a corresponding bit $v_R[i] = 1$ must be retrieved from disk and examined to ensure they meet the selection criteria. All records $r_i$ with a corresponding bit  $v_R[i] = 0$ are \textit{pruned} immediately and do not require retrieval from disk. Because a well-designed bitmap is sparse and compressible, it can be stored in core memory, which is orders of magnitude faster than disk.

As such, bitmaps help reduce disk accesses when properly discretized, resulting in a space/accuracy tradeoff. More precise pruning may have been possible had we split the attribute $Y$ into even finer-grained bins. However, each additional bin effectively adds an entire dimension, increasing the bitmap index size, thereby challenging its ability to fit in core memory.

\subsection{Monitoring and Live Analysis}
% \todo[inline]{dc: hmm not sure what others think, but this content was not what I was expecting to read about the Monitoring and Live Analysis subsystem. It talks more about the data set that we got from BPA. What I was expecting was a high-level overview of what event signatures were, what the correlation matrix is doing, and so on.}

%\todo[inline]{dc: I'm not sure if this subsection is even needed. The details on the PMU data also seems out of place. I think it belongs in the Results section.}
The Monitoring and Live Analysis subsystem contains our implementation of an event detection system. As with the Historical Data Management System, a different event detection algorithm can be used in the correlation matrix's place.

A challenge with the increasing deployment of PMUs in power systems is the large amounts of data from those sensors. So far, pre-processing methodologies to handle high-cardinality data from PMUs are not widely available, and little progress has been made to streamline and consolidate these algorithms. Therefore, the two major capabilities necessary to maintain interoperability between raw power system data and our correlation methodology are described below -- namely data playback and data storage.

\begin{figure}[ht]
\centering
\includegraphics[trim=0 0 0 0, clip, width=.85\textwidth]{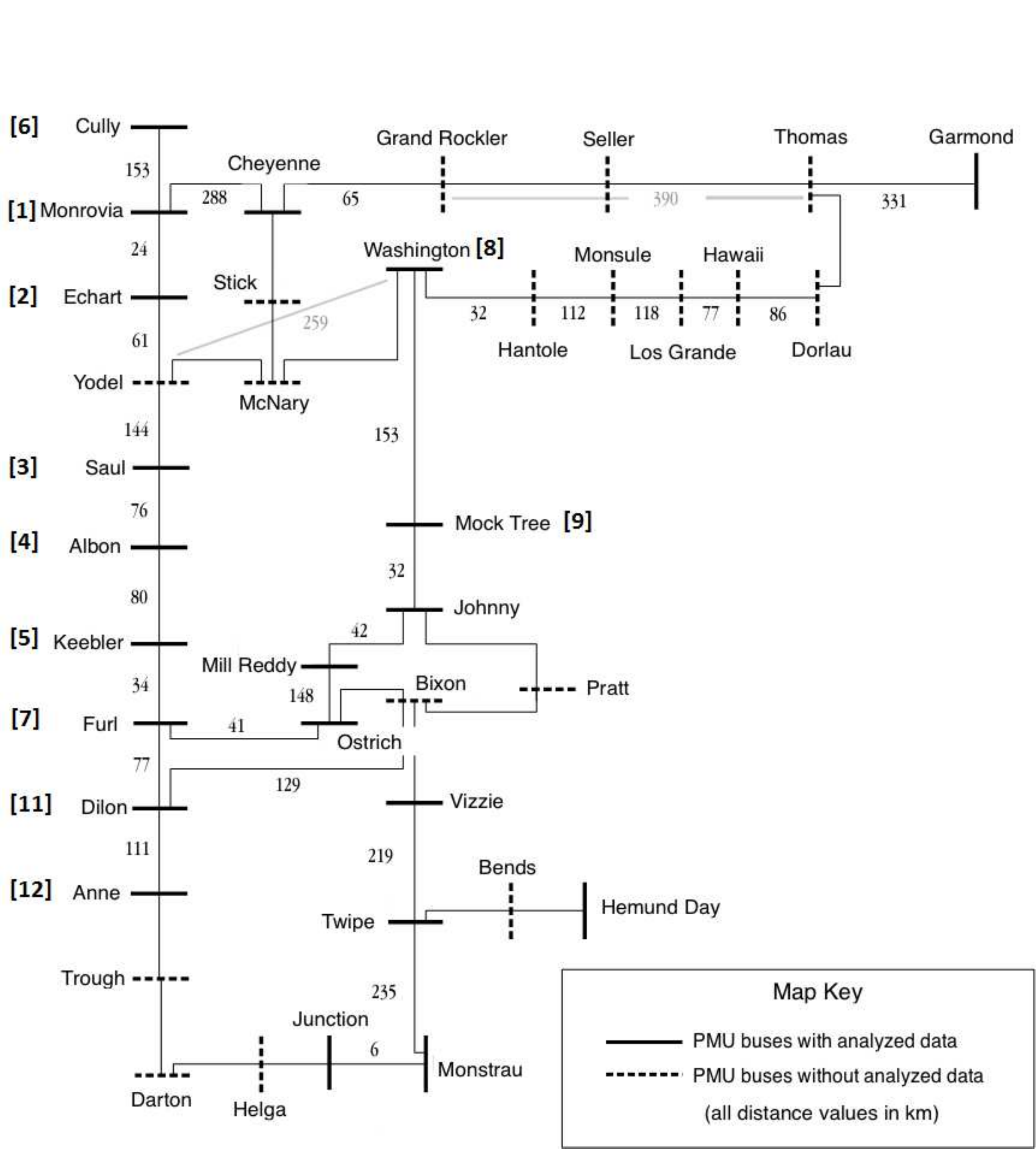}
\caption{The relative location of and distances between PMU sites in kilometers (not to scale).  Note the location of the Monrovia bus, upper left, which serves as a test case in our Results section. Bracketed numbers correspond to correlation visuals in Results section.}
\label{fig1}
\end{figure}

The one-year data set we assessed includes information from August 2012 to August 2013. It totals 950 GB of positive-sequence voltage magnitude ($V$) and positive sequence voltage phase angle ($\phi$).  Each measurement is represented by a \textit{date/time} and its corresponding \textit{phasor} value.  These measurements are acquired every $0.0167$ seconds ($60$ Hz). The phase angle $\phi$ is a time-varying real number that oscillates within the range of $[-180,180]$. The voltage, on the other hand, is a non-negative real number. Each file in the set typically holds one to five minutes of data from each of the 20 separate PMU sites. We opted to consolidate and standardize these files for ease of input into our PDC engine. Once the data is coalesced, it is fed into our event detection algorithm described in Section ~\ref{sec:methodology}.

\section{Methodology}
\label{sec:methodology}

\subsection{Data Input \& Correlation}
\label{sec:DataInputAndCorrelation}
%\todo[inline]{(BM) Optimize correlation for online data. This includes, but is not limited to, further research onto varying window sizes, different data stream rates (currently only being run at 60Hz, we should test higher and lower frequencies)\ldots not to sure what else here.}
As positive sequence voltage data is generated in the time-domain by our PDC, the data must be read into the working memory of our correlation algorithm. In an effort to minimize computational complexity, we developed a custom data structure in order to quickly append new data, reference data already stored, and account for multiple characteristics such as time, magnitude, phase, and correlation coefficients, for each of the 20 PMUs.

Besides pre-processing PMU data, a key aspect of a decision-making framework is to accurately identify events during a real contingency situation.
%the main contribution of this research is to provide a method that can be used for decision-making during grid operation, especially during contingency situations. 
In order to achieve this level of operator support, however, we must be able to distinguish between data errors and power system events.  

We propose a correlation technique that can be used to flag specific data and power system events.Our algorithm calculates the correlation coefficient between parameters at different substations. Consider, under normal operating conditions, electrical paramters measured at one substation will be very similar to those measured at an adjacent substation due to close electrical proximity.   As such, the correlation coefficients between parameters at different substations will be very near one.  The parameters measured by PMUs include the magnitudes and phase angles of phase voltage and line current, the magnitudes and phase angles of voltage and current sequence components, frequency, and rate of change of frequency (or ROCOF).

During an event, such as a power system transient or a data error, the measurements of a particular parameter at two different substations will differ, at least temporarily. And so, the correlation coefficients will deviate away from one during the event.  For demonstrative purposes in this paper, we use the positive sequence voltages.  Consider a lighting event near substation A, which affects the positive sequence magnitude at that substation.  The lightning event will also affect the positive sequence voltage at adjacent substation B, though with a lower magnitude and a time delay due to the intervening line reactance.  Our correlation algorithm would detect a deviation between the positive sequence voltage magnitudes at these two substations, returning a correlation coefficient less than one during the event.

Our algorithm simultaneously calculates the correlation coefficients between more than two substations, and in fact between more than one parameter.  As such, we get upper triangles of size $N^2$ of correlation coefficients, allowing us to monitor the correlation of parameters between a suite of substations. Below, we expand on the correlation methodology that was developed in order to identify events.
%Correlation is a well known mathematical and statistical method for determining the compatibility of large data sets. Specifically, the Pearson Product-Moment correlation determines how well data is linearly correlated, and has been used successfully in other graph-based problems such as ~\cite{Paper041}.

The correlation detection algorithm need not scale with $N^2$ as more PMUs are added to a balancing area.  In our current work, we demonstrate that the algorithm need only analyze a handful, 4 or 5, of real-time PMU data streams concurrently to reliable detect local events. We envision the algorithm would be hosted on phasor data concentrators (PDC), which aggregate PMU data streams, and these PDCs would be widely distributed throughout a balancing area.  Each PDC would host an instance of the algorithm for detecting events within its immediate vicinity.  As such, the algorithm would not face an $N^2$ issue as more PMUs are added to a balancing area.

We start with a formal definition of the Pearson Correlation index. Given two independent input sets of data $X$ and $Y$ of length $N$ ($X$ and $Y$ being either the momentary magnitude or phase-data values of two PMU site readings), we obtain a correlation coefficient $r$ between $-1$ and $1$ based on the following equation:

\begin{gather*}
\label{eq1}
r = \frac{\Sigma{(XY)}-\frac{\Sigma{X}\Sigma{Y}}{N}}{\sqrt{(\Sigma{(X^2)} - \frac{(\Sigma{X})^2}{N})\times(\Sigma{(Y^2)}-\frac{(\Sigma{Y})^2}{N})}}
\end{gather*} 

Two modifications and application-specific improvements were made to this mathematical formula. First, the algorithm was made incremental. In this way, each data point could be read in from the PDC feeder and immediately incorporated into its correlation coefficient without the need to directly calculate each summation, average, and standard deviation repeatedly at each time step. 
% This helps reduce run time.
Second, we maintain correlation information over varying windows of time. We used a  queue to keep  separate pointers to end positions of each defined sliding window. 
%This is seen clearly in Figure~\ref{fig3}. It is worth noting here that separate lists for each sliding window are \textit{not} created. Rather, pointers in a single list are maintained to minimize memory usage, as well as minimize copies of data already managed. 

%\begin{figure}[!htp]
%\centering
%\includegraphics[width=.4\textwidth]{images/SlidingWindowImage}
%\caption{The window sizing for data queue.}
%\label{fig3}
%\end{figure}

The addition of this multi-window-size feature allows for pairs of PMUs to be correlated over different time intervals concurrently. This design allows different events to be identified based on different sliding window sizes. This capability to correlate over multiple discrete periods of time is especially useful in determining if suspect correlations are due to data issues, or are in fact real disturbances. In our approach, large window sizes correspond to 1200, 600, and 60 data points (20 sec, 10 sec, and 1 sec, respectively). We use smaller, multi-cycle window lengths (54, 48, 30, 18, 12, and 6 data points) to assist with identifying the difference between data events and power system contingencies.  Data events are readily detected with those short window lengths, as data errors cause rapid decorrelation between PMUs.  Power system events are detectable using longer window lengths, depending on the type of event.  We hypothesize that fast events such as lightning strikes are detectable using moderate-length windows (1 to 10 seconds) while detection of slow events, such as inter-area oscillations, would require longer window lengths. The distinction here is important because, with any large-scale data set, there is a question of data validity. It is of strategic importance to identify false data originating from PMU inaccuracies, especially since these devices are used to inform higher-level applications such as state-space estimation and remedial action schema.

\subsection{Bitmap Engine}

Given a user query that selects a subset of records from the PMU data archive, the na\"{i}ve approach to respond to the query would be to perform a linear scan of the database, comparing each record for selection, and then returning the matching records. For a real-time application such as power system situational awareness, this operation would be too expensive because disk I/O operations are slow. Our PMU data management system has multiple software components that allow a user to build a bitmap index over raw data, and to efficiently query records that match specifications.

%Our PMU data management system, depicted in Fig.~\ref{figure:sysdesign}, has multiple software components allowing the user to build a bitmap index over raw data, and to efficiently query records that match specifications.

% we must reduce the total query time by reducing the amount of I/O.

The \textit{Bitmap Creator} inputs the raw PMU data and generates a bitmap using the binning strategy specified below. When new files are added to the database, these records are appended onto the index. Once the bitmap is created, the \textit{Compressor} will compress the index using WAH. After compression, the system is ready to receive queries from the user. These queries will give selection conditions on which values of particular attributes the user is interested in. The \textit{Query Engine} then translates the query into boolean operations over the specified bins in the compressed index. This then produces a \textit{Result Bit Vector} $v_R$ that contains information on which records need to be retrieved from disk. 
% \begin{figure}[htp]
% \centering
% \includegraphics[width=0.4\textwidth]{eps_images/BitmapSystemDesign.eps}
% \caption {PMU Data Management System Architecture}
% \label{figure:sysdesign}
% \vspace{-2em}
% \end{figure}

While $v_R$ holds the selected record information (all bits with a value of 1), it is the actual data on disk that must be returned. An intermediate data structure, the \textit{File Map}, was created to facilitate this role.
The File Map is an intermediate data structure that holds metadata on the files and how many tuples \footnote{An entry (or record) in the database} they each contain. There are two values per File Map entry: \textit{totalRowCount} and \textit{filePointer}. The \textit{totalRowCount} contains the total number of tuples up to and including that particular file. The \textit{filePointer} holds a pointer to the corresponding file on disk that contains the next set of tuples. To retrieve files with this method, the result bit vector is first scanned and a \textit{count} is kept for the number of bits that have been read. For each hit, the count is hashed to its corresponding index in the File Map. This is an upper-bound hash, meaning that the count value is hashed to the closest \textit{totalRowCount} value, without being greater than it. This will give the corresponding file that is desired.

\begin{figure}[ht]
\centering
\includegraphics[width=0.85\textwidth]{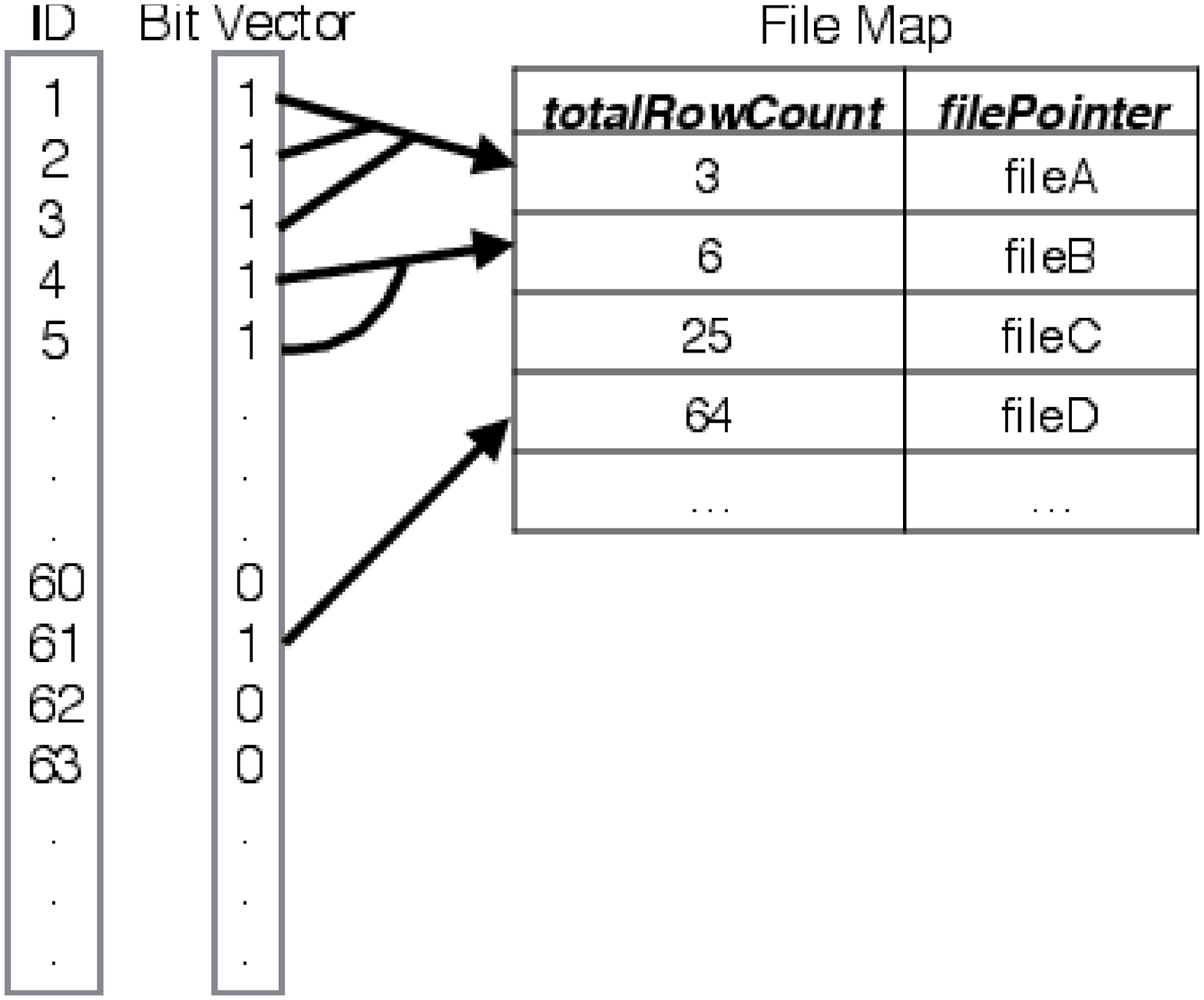}
\caption {File Map Structure}
\label{figure:filemap}
\end{figure}

Fig.~\ref{figure:filemap} illustrates a small example of a bit vector and where the bits hash to the filemap. Bits one through three are hashed to the first row in the File Map structure. Bits 4 and 5 are hashed to the second row since these bits represent tuples 4 and 5, which are stored in \textit{fileB}. With the upper bound hash, bits 4 and 5 hash to \textit{totalRowCount} 6, since they are both greater than 3 but less than or equal to 6. Bits 60, 62, and 63 are not hashed since they are not hits. Only bits in the bit vector that have value one will be hashed. This leads to improved performance when there are long stretches of zeroes in the bit vector.

\subsection{Binning Strategies}

We obtained data from $20$ PMUs within Bonneville Power Administration's (BPA) balancing area from August 2012 to August 2013. At each PMU, a \textit{phasor} measurement is sampled every $1/60$ sec. Each measurement is represented by a \textit{date-time} and a \textit{phasor}, which is a pair of values: the phase angle $\phi$ and the positive voltage magnitude $V$. The phasors from the $20$ PMUs are combined, resulting in $2 \times 20~\text{PMUs} = 40$ attributes.
The phase angle $\phi$ is a time-varying real number that oscillates within the range of $[-180,180]$. The voltage, on the other hand, is a non-negative real number. In order to define the bitmap ranges, we examined $\phi$ and $V$'s distributions. We analyzed the distribution of $\phi$ and $V$ over a sample size of $30$ days ($155,520,000$ measurements). 

% The data set is split into multiple flat files, with each file storing five minutes measurements. This results in $18,000$ rows per file. \TODO{DC}{Check! In all, there are a total of $XX$ measurements.}
% and present the cumulative distribution function (CDF) of $\phi$ and $V$ in Figures \ref{} and \ref{}, respectively. 

% The bitmap was designed with the notion that the majority of queries over the history data set would correspond to certain dates/time and events in the system.

To optimize for speed, the design of the bitmap must be informed by the queries that will be frequently executed. For frequently queried values in bitmap structures, a crippling factor in response time is the candidacy checks to identify true positives, which require disk access. Due to imperfect discretization, bins will often contain bits that indicate more than one value. It is therefore necessary to check whether that bit is an indication of the correct value. For example, if a bin has the range of five possible values then that means each bit in that bin is one of five different values. Performing this check, called a \textit{candidacy check}, ensures that the \textit{tuple} contains the desired value for the query. Choosing the correct binning strategy can therefore potentially improve our query times by reducing candidacy checks among values that were expected to be queried.

From discussions with power systems experts at BPA, queries typically comprise a specific range of dates, voltage $V$, phase angle $\phi$, or any combination of these attributes. When generating the bitmap, the binning (discretization) strategy can minimize candidate record checks and provide fast query response times. Due to the low cardinality of the \textit{date-time} attribute, it was simple to generate bins: 60 bins each for second and minute, 24 bins for hour, 31 bins for day, etc. with the exception of the year. In this case we used 11 bins for the year, starting at 2010. Since there were no range bins, no candidacy checks were necessary when performing queries on the dates. Because $\phi$ and $V$ are real values, we discretize based on their distribution. In order to find the distributions of both $\phi$ and $V$, the cumulative distribution function (CDF) plots were constructed. These distributions determined what binning strategies were used.

% The cardinality plays a large role in how the attributes should be binned when creating the bitmap index. 
% The date attribute has a very low cardinality because the data has a start and end date making it contain values of a finite range. Because of this it is possible to have exact binning for all values of the Date attribute, except milliseconds. This required small bins due to its range of values. 

\begin{figure}[ht]
\centering
\includegraphics[trim=0 0 0 0, clip, width=0.85\textwidth]{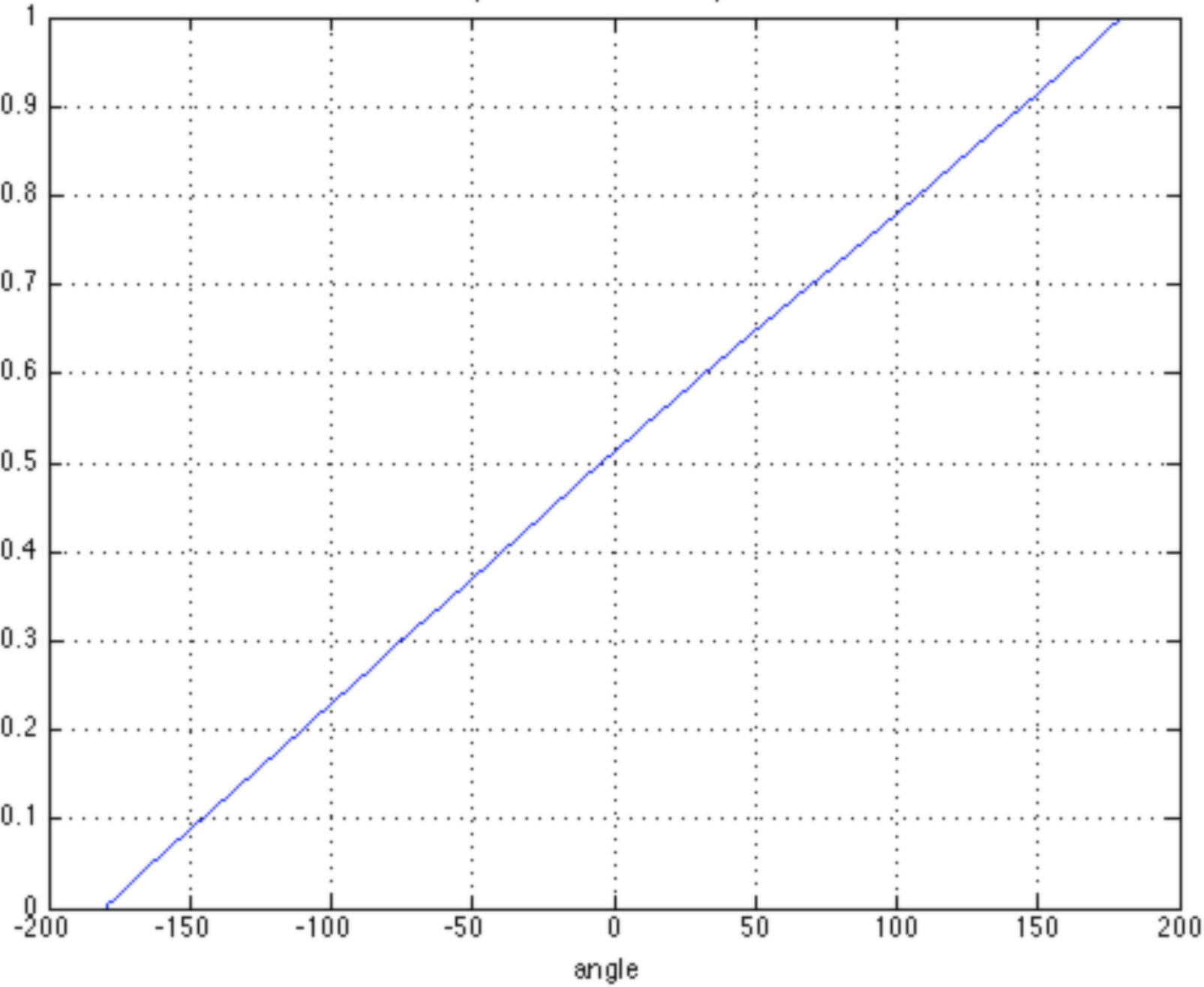}
\caption{Normal Phase Angle CDF}
\label{fig:cdfphase}
%\end{figure}

%\todo[inline]{RB: Not for this paper, but we should perhaps be looking at CDFs of phase angle \textit{differences}, though this would lead to a sparse $N^2/2$ data set for phase angle information.}

%\begin{figure}[ht]
%\centering
\includegraphics[trim=0 0 0 0, clip, width=0.85\textwidth]{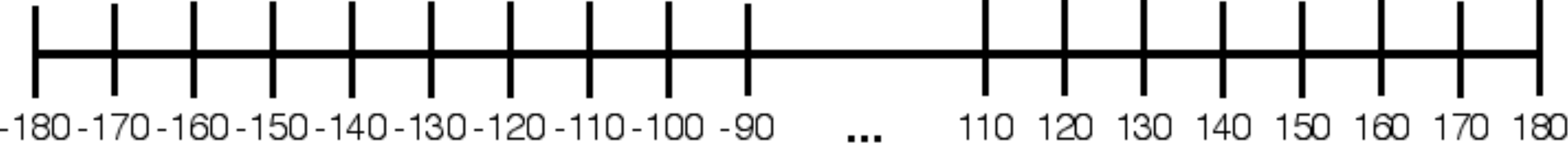}
\caption{Phase Angle Bins}
\label{fig:binphase}
\end{figure}

Fig.~\ref{fig:cdfphase} illustrates the phase angle $\phi$ distribution. From this graph we can see that $\phi$ follows a uniform distribution. Because $\phi$ is also bounded, we apply an \textit{equal-width} binning strategy over $\phi$, meaning the range of each bin is  equivalent. We designed the \textit{bitmap creator} in such a way that this range can be assigned by the user before creation of the bitmap. For our experiments we set this value to 10, leaving 36 bins for each PMU attribute. Fig.~\ref{fig:binphase} represents the phase angle values that were assigned to each bin.

\begin{figure}[ht]
\centering
\includegraphics[trim=0 0 0 0, clip, width=0.85\textwidth]{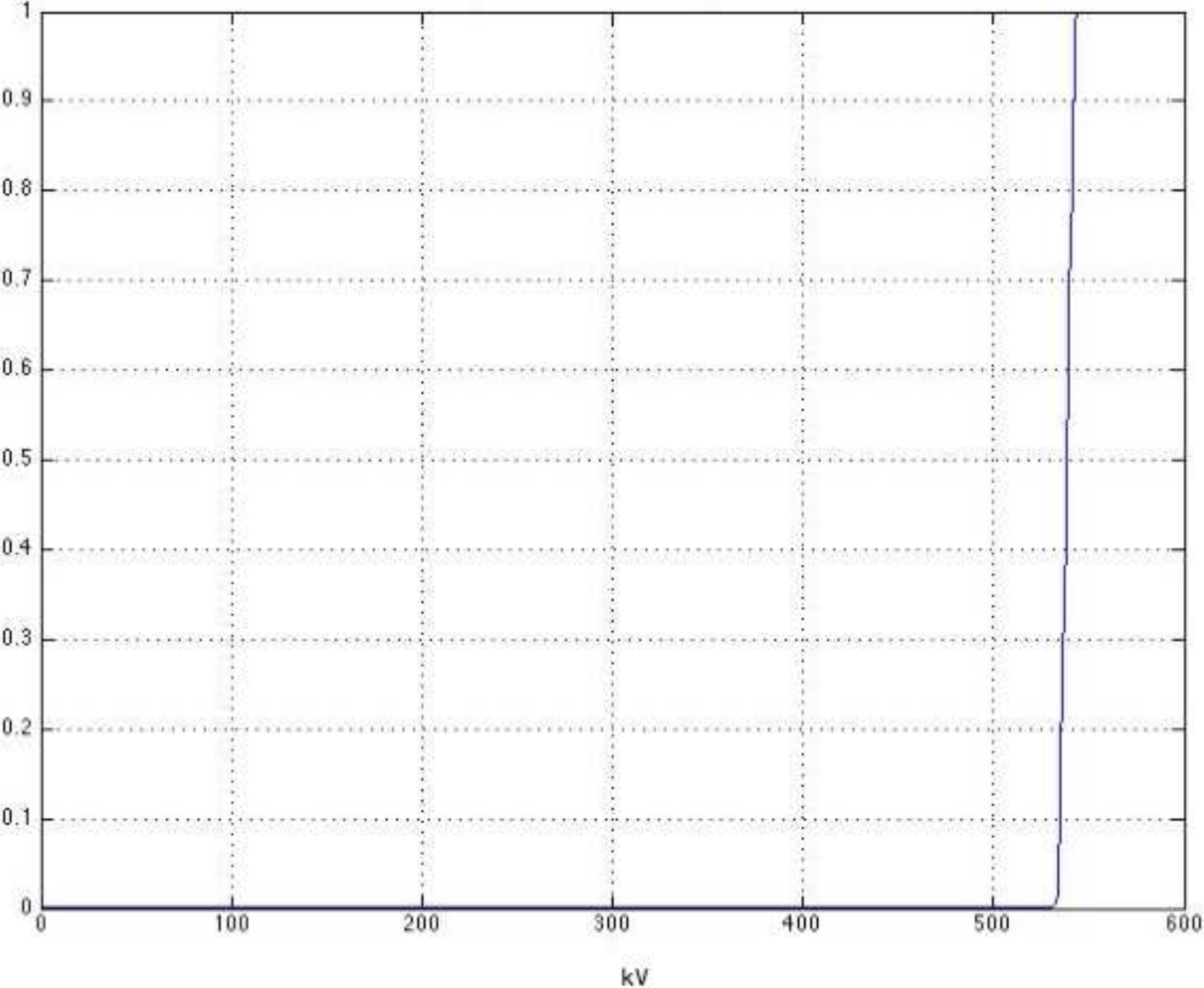}
\caption{Normal positive sequence voltage magnitude CDF}
\label{fig:cdfmag}
%\end{figure}

%\begin{figure}[ht]
%\centering
\includegraphics[trim=0 0 0 0, clip, width=0.85\textwidth]{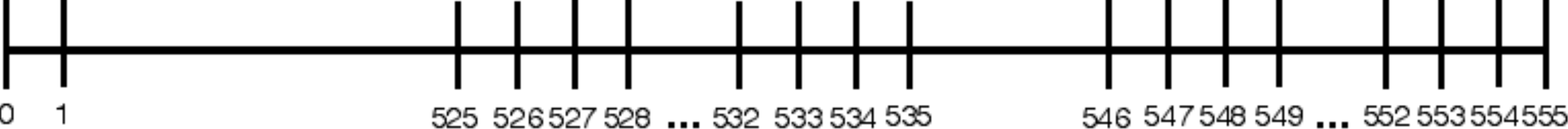}
\caption{Voltage Magnitude Binning}
\label{fig:binmag}
\end{figure}

Fig.~\ref{fig:cdfmag} illustrates the distribution for normal operation of a PMU's voltage magnitude. The data set only contains positive sequence voltage. The majority of the values occur between $[535,545]$. For this attribute, we used a binning strategy which attempts to minimize candidacy checks for the values that are most likely to be queried. We assume the majority of queries from the user will pertain to some anomaly, that is values that are not apart of normal operations. Therefore, a bin with range $[535,545]$ can be created to contain the regularly occurring values. Since the range of the bin is quite large, and it spans the values which occur most frequently, then the majority of tuples that fall into this category will require candidacy checks. However, our assumption is that queries will occur for abnormal values. This leads to a specific strategy for binning: There are ten bins on either side of the central bin which represents the normal operational range. Each of these outer bins is capable of containing a value with a range of one. Fig.~\ref{fig:binmag} represents the binning distribution for voltage magnitude. There is an additional bin for the value zero, since this is an indication of a data event at a PMU site. This strategy generates bins of small ranges for values of $V$ that will be queried frequently and very large bins for those that aren't.

Currently the PMUs that we are utilizing do not report measurements in per unit. To avoid adding another layer of computation, the measurements are binned according to their physical units. These binning strategies are applicable to other PMU networks with different nominal voltages, or one could decide to adopt the per unit system in the first layer for a specific implementation of this framework. 

In addition to the aforementioned attributes, we also introduced an attribute $\Delta$, which represents the displacement between phase angles from the previous time-stamp, i.e., $\Delta_t = |\phi_t - \phi_{t-1}|$.  $\Delta$ is a coarse representation of rate of change and can be an indicator as to whether a power event occurred. Therefore, we bin $\Delta$ with smaller ranges, reducing the number of candidacy checks.  Listed in Table ~\ref{table:bins} are the number of bins that we used for each attribute. The total is 4,988 bins for each row in the bitmap index.
\begin{table}[ht]
\begin{center}
\begin{tabular}{|c|c|c|c|}
\hline
Attr. & \# Bins & Attr. & \# Bins \\
 \hline
Year & 11 & Month & 12 \\
 \hline
Day & 31 & Hour & 24 \\
 \hline
Min. & 60 & Sec. & 60 \\
 \hline
mSec. & 10 & $\Phi$ & $20\times 23$ \\
 \hline
$V$ & $20\times 36$ & $\Delta$ & $20\times 180$ \\
\hline
\end{tabular}
\end{center}
\caption{\label{table:bins}Bins}
\end{table}

To demonstrate how well the bitmap can scale and compress the data, 4200000 million tuples of our database have been compressed using WAH with 32-bit words. The original size of the bitmap index was ~2.75 GB. Once compressed, the bitmap index was 8.03216 MB. This means this bitmap index has a compression ratio of 342.37, $\frac{uncompressed}{compressed}$. Our query engine model scales more efficiently than other commonly used querying engines. For this paper, we used MySQL as a comparison. MySQL does not perform any compression on it's index, therefore it will require more space and when the tuple count increases, it may not even be able to fit into memory. MySQL also performs significantly worse on datasets with a very high number of tuples in a single table.

% \begin{itemize}
% \item Year: 11
% \item Month: 12 
% \item Day: 31
% \item Hour: 24
% \item Minute: 60
% \item Second: 60
% \item Millisecond: 10
% \item $\phi$ (23 for each PMU): $20\times 36$
% \item $V$ (36 for each PMU): $20 \times 23$
% \item $\Delta$ (180 for each PMU): $20 \times 180$
% \end{itemize}

\section{Results}
\label{sec:results}
%Not too sure how to subsection this

This section focuses on highlighting some of the preliminary qualitative information obtained by processing and analyzing the PMU data streams via our correlation methodology as well as quantifying the query times run on the database. This consists of visualizing the PMU data for a particular case study at the ``Monrovia" bus seen in Fig.~\ref{fig1}. Two types of queries were run, linear scan and bitmap indexing, which are compared in Table ~\ref{table:querycomp}.

% Removed(RM)  We begin with a short introduction to our visualization technique and the analysis behind producing it. Next we follow up with a case study at one particular bus in the system (depicted in Fig.~\ref{fig1} and Fig.~\ref{EventMap}) -- namely the Monrovia Bus. We demonstrate that the algorithm is capable of identifying the difference between data events and power system events. 

%\todo[inline, color=red]{Change figure 13 such that it only includes the PMUs referenced in figure 15. Maintain the topological distance ordering though. JL: Done.}

\subsection{Visualization Structure}

The purpose of this subsection is to introduce the layout of the visualization structure used in the case study in Sec.~\ref{casestudy}. First, each coordinate (square) represents the correlation coefficient of the two PMUs that make up its coordinates. The color of the square represents how close the correlation is to $1$ or $-1$, and the sign at the coordinate represents either positively correlated or inversely correlated PMU pairs. Typically a magnitude of correlation above $0.4-0.5$ is considered correlated. Thus any squares depicting blue shades would be considered de-correlated. It is important to keep in mind that this visualization is temporal, and represents different time window lengths as discussed in Section \ref{sec:methodology}.
%(Fig~\ref{fig3}).

% REMOVED (RM) 
%\begin{figure}[t]
%\centering
%\includegraphics[width = \columnwidth]{images/Mag600_0010_0000.png}
%\caption{Example PMU correlation visualization (topological distance) over 600 data points (10 seconds).}
%\label{exampleVis}
%\end{figure}

% REMOVED (RM) (this is exlained in the case study writing... ) The blacked out columns and rows with the null symbol represent PMUs that were offline during that particular time. This is typically given by the PMU node producing $0$ data. Our algorithm detects this, and in the later case study we remove these PMUs because they are consistently yielding unusable data.

Next, this visualization incorporates electrical distance into the spatial organization of each monitored bus. The notion of electrical distance has been proven useful in multiple power systems applications, but was developed most notably by Cotilla-Sanchez et al.~in \cite{Paper031} for the purpose of multi-objective power network partitioning. In our data set, adjacent cells within the triangular visualization matrix are referenced to PMU 1, either topologically, or electrically. We anticipate this organization of PMUs will produce electrically coherent zones. As a result, the visualization will naturally cluster, thus benefiting ease of analysis and application of advanced techniques such as pattern recognition.

%% Begin Jordan's Data -- Case Study

\subsection{Monrovia Event Case Study}
\label{casestudy}
In order to demonstrate some preliminary identification of data and power system events, we analyzed a subset of contingencies that were known to have occurred at the Monrovia Bus. We address PMU data drop and PMU data misread contingencies, as well as a known lightning event near the Monrovia bus.

%\subsubsection{Clean Data}
 
%\begin{figure}[t]
%\centering
%\includegraphics[width=.4\textwidth]{images/Normal_Op_10s_Prior.png}
%\caption{Voltage magnitude plot of the Monrovia bus during nominal operating conditions.}
%\label{cleanMAG}
%\end{figure}
%\begin{figure}[!h]
%\centering
%\includegraphics[width=.4\textwidth]{images/Mag600_0009_prior.png}
%\caption{Visual depiction of the correlation values during nominal operating conditions (electrical distance), taken over a 10 second window.}
%\label{cleanVIS}
%\end{figure}

% cut to remove fig 11.........Figure~\ref{cleanMAG} depicts a single bus's positive sequence voltage magnitude plot over a ten second window. It is worth noting that the y-axis scale is zoomed in tightly around $1.081$ [PU]. 
%Data sets void of any data issues or disturbances are categorized as `clean' data sets. We know (from archived notes and comments) that these subsets of data exhibit the system operating under normal circumstances. Next, we characterized correlation when using a ten second window. Figure~\ref{cleanVIS} depicts the correlation coefficients for every combination of PMUs in the network during nominal operation. The correlation matrix depicts how well coupled each PMU is to every other. As expected, we see that near-by buses are strongly correlated within distinct clusters, though there is some noise associated. 

\subsubsection{PMU Data Events at Monrovia}

\begin{figure}[ht]
\centering
\includegraphics[width=.85\textwidth]{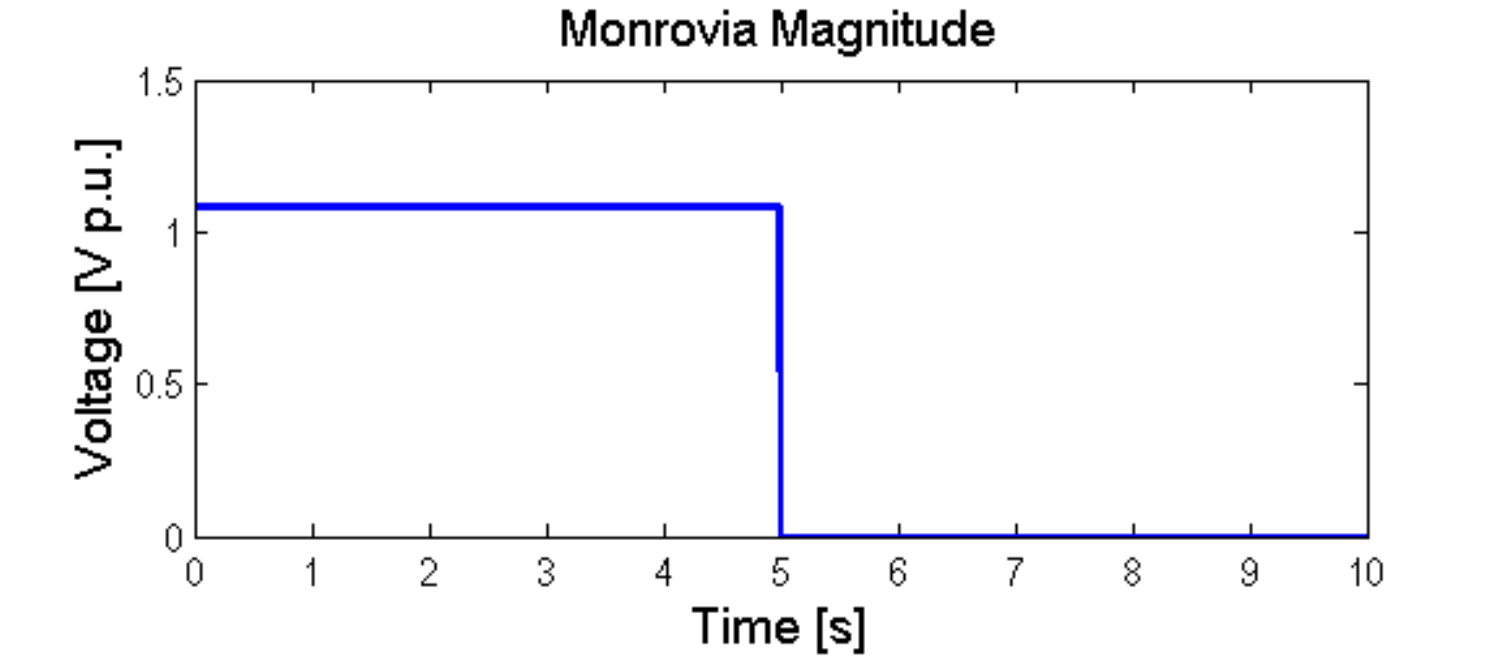}
\includegraphics[width=.85\textwidth]{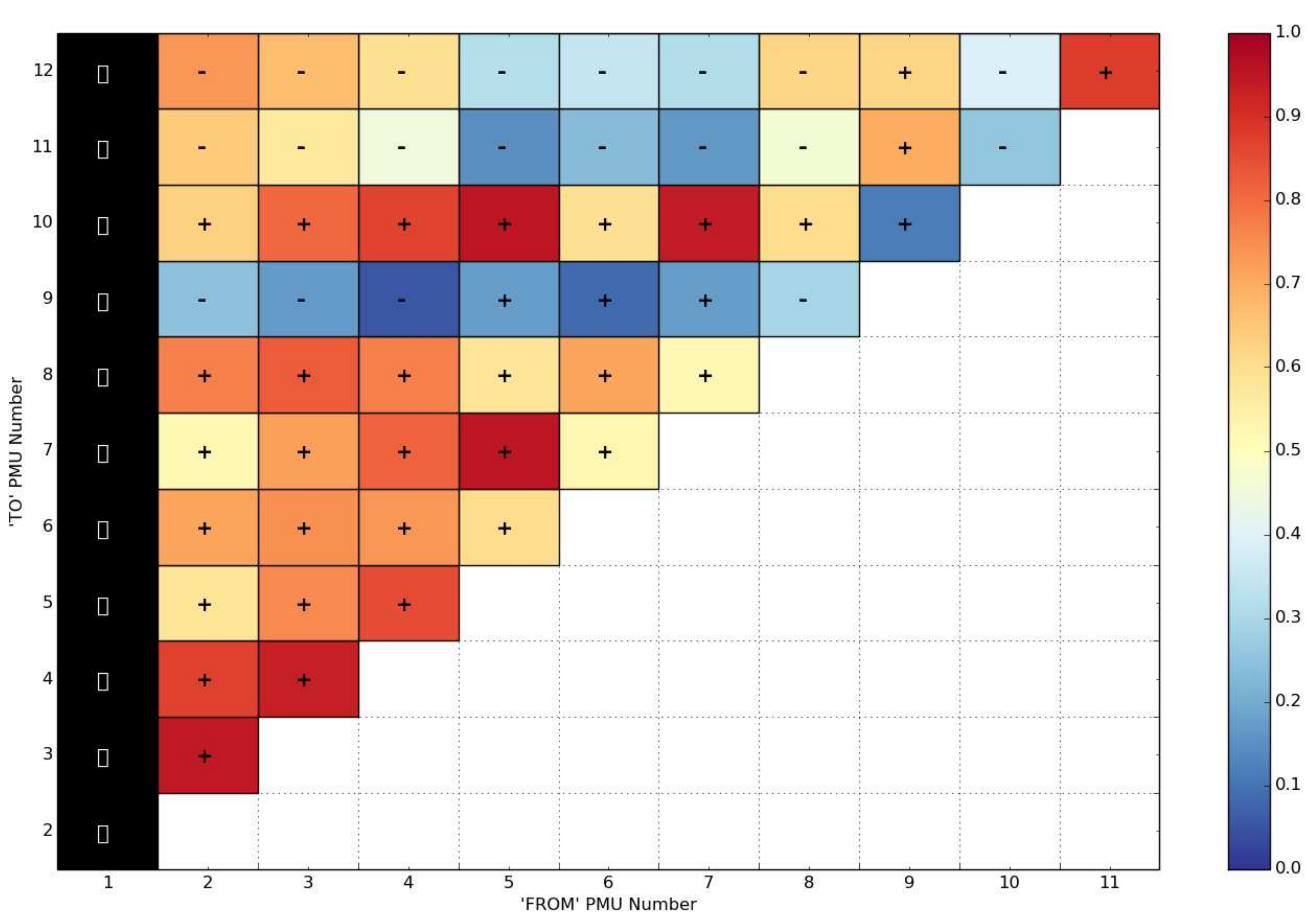}
\caption{A flagged ``Data drop" event at the Monrovia bus with $\frac{1}{10}$ sec. sliding window (electrical distance).} % is (electrical distance) needed?
\label{datadrop}
\end{figure}

\begin{figure}[ht]
\centering
\includegraphics[width=.85\textwidth]{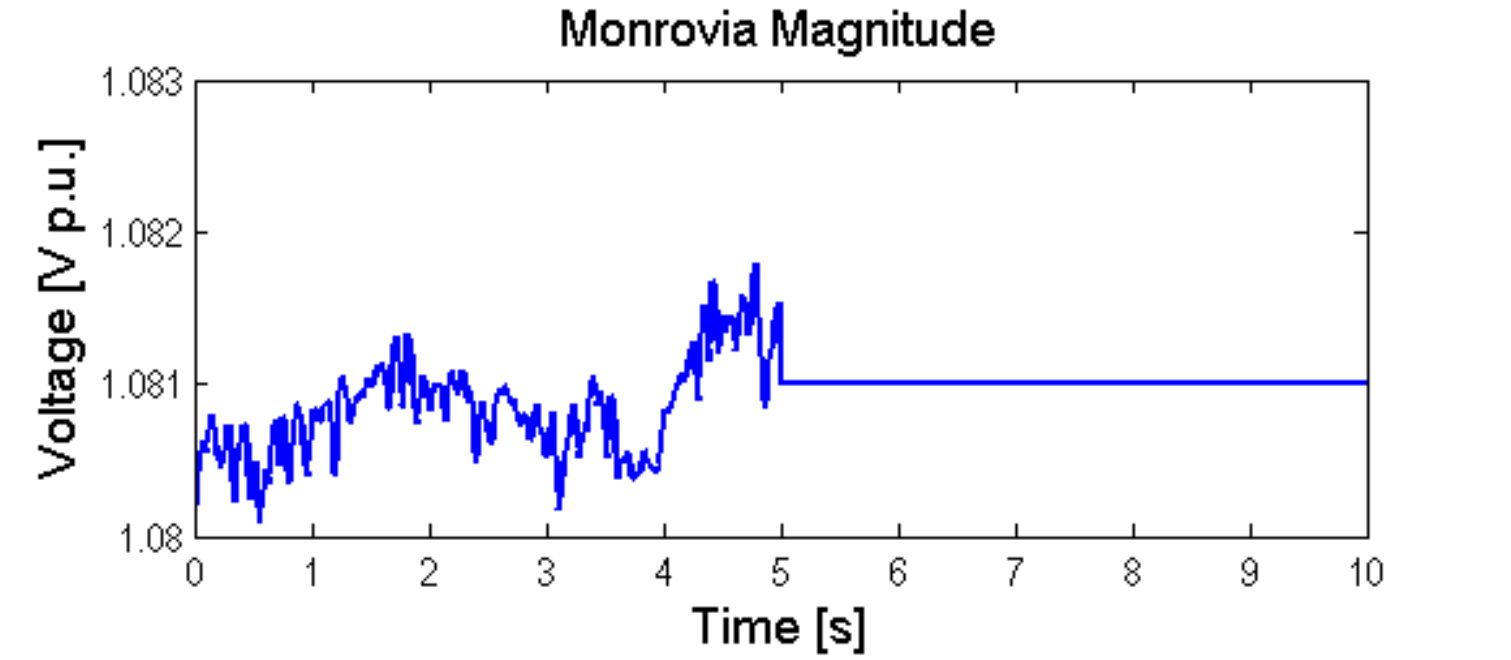}
\includegraphics[width=.85\textwidth]{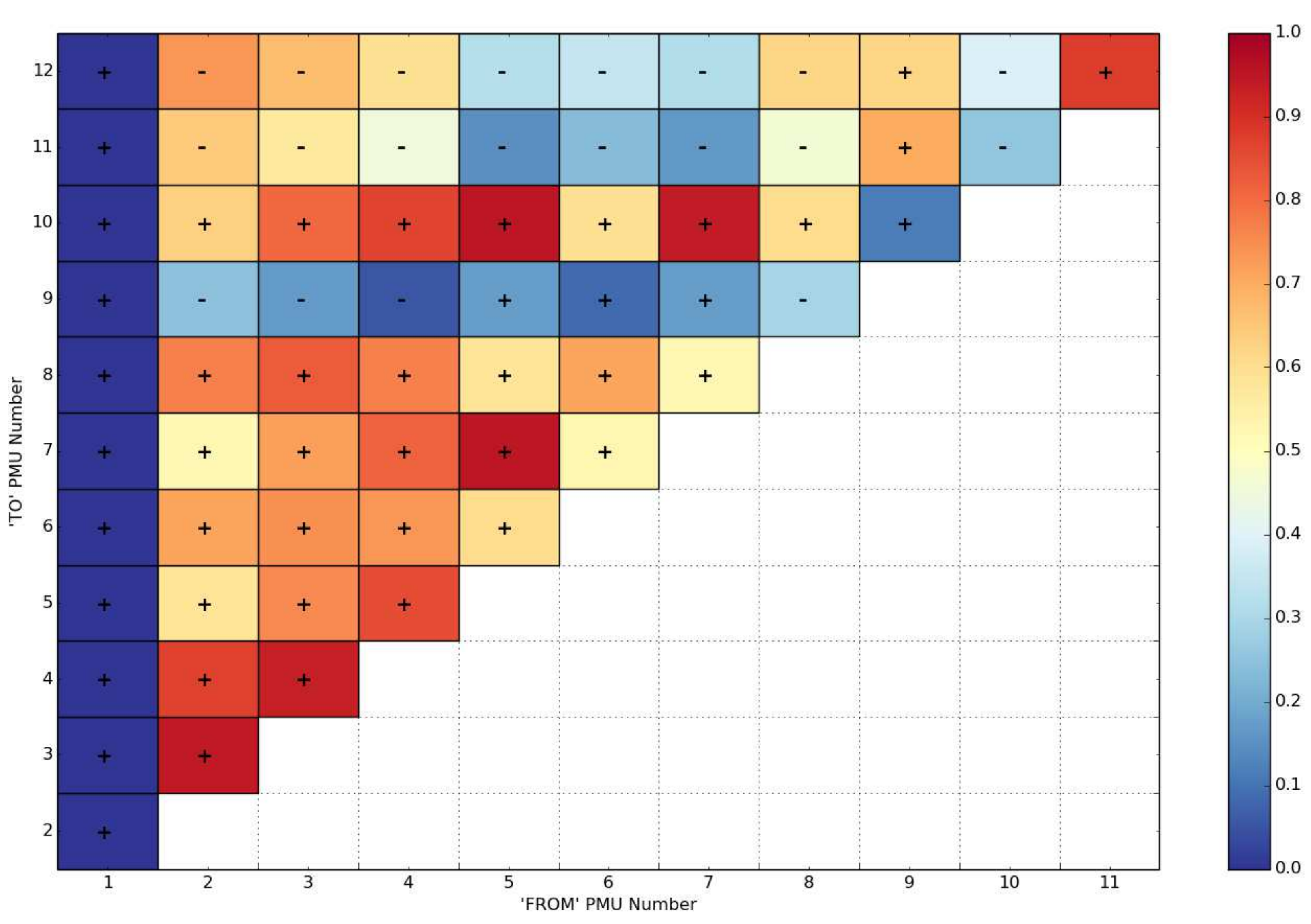}
\caption{A flagged ``PMU Misread" event near the Monrovia bus with $\frac{1}{10}$ sec. sliding window (electrical distance).}  % is (electrical distance) needed?
\label{misread}
\end{figure}

PMU data streams must be validated as accurate in order to ensure reliable and effective decisions are made during grid operation. Invalid data can be introduced in multiple ways, and so far we have created the ability to detect two specific data events. First, the PMU may go offline resulting in a constant stream of ``zero" data (termed ``data drop") as seen in Figure~\ref{datadrop}. Second, a PMU data stream may produce unreliable data, which is characterized by repeatedly producing the same measurement over a discrete window of time, as shown in Fig.~\ref{misread}. 

% This had to be cooked up such that the whole 10 sec time frame was zeroed out. (RM-- No problem, we know this is how it acts anyway. Just that the correlation needs the whole window to be zeroed. This bit of fudging is still accurate.)

Both of these data contingencies are flagged by our algorithm using the small window sizes, as typical data events occur in sub-second time frames. As seen in the images, the full-column pattern of null data (blacked out column) and the severe de-correlation both indicate a data event at the Monrovia Bus.

\subsubsection{Power System Event at Monrovia}

\begin{figure}[ht]
\centering
\includegraphics[width=.85\textwidth]{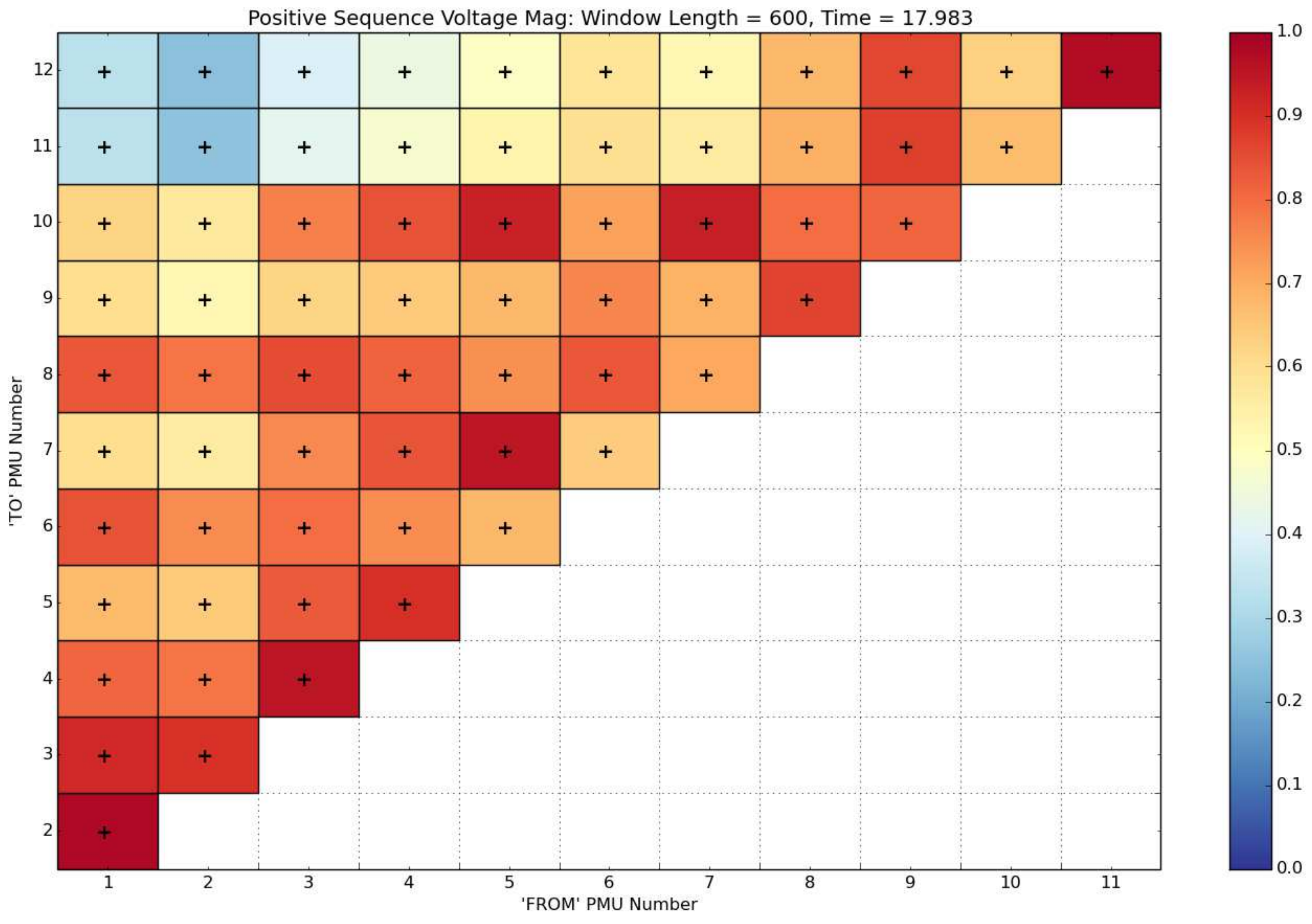}
\caption{Monrovia lightning event correlation over 10 sec. sliding window (electrical distance).}
\label{weak600}
\end{figure}

The final type of event that our correlation technique is currently able to characterize is when a power system lightning contingency occurs. Again, for this case study, we focus on a known lightning event at the Monrovia bus. For this particular lighting strike, we run the correlation algorithm over a window size of 10 seconds. The visualized results can be seen in Figure~\ref{weak600}.

%REMOVED ALL THIS AND REWORDED (RM) 
%As will be discussed in Section~\ref{sec:future} the ultimate goal is to flag specific events outlined in Table~\ref{table1}, however for this case study we focus on a known lightning event near the Monrovia bus. %(indicated in Fig.~\ref{EventMap}).
%Figures~\ref{largeWIN} and \ref{smallWIN} depict the lightning event in the time domain. In this section we run the correlation algorithm, on 

%REMOVED (RM) can be seen in background map.
%\begin{figure}[t]
%\centering
%\includegraphics[width=3in]{images/PMU_Mapping_Rev_4.png}
%\caption{Lightning event at Bus Monrovia. Bus number assignments align with correlation visualizations (electrical distance)}
%\label{EventMap}
%\end{figure}

%\begin{figure}[t]
%\centering
%\includegraphics[width=.4\textwidth]{images/Mag6_0009_lightningStrike.png}
%\caption{Monrovia lightning contingency over a $\frac{1}{10}$ sec. sliding window (electrical distance).}
%\label{weak6}
%\end{figure}

%There are a couple of deductions that can be made from these power system contingency correlations. First, it is feasible to identify clusters of PMUs that change drastically from highly correlated (as in Fig.~\ref{cleanVIS}) to highly uncorrelated (as seen in Fig~\ref{weak600}). This suggests that it would be possible to use higher order classification methods or pattern recognition to identify this type of contingency. Second, inspecting Figure~\ref{weak600} and referring to Figure~\ref{fig1}, we see emergence of how electrical distance influences correlation. Since buses south of Monrovia are along a parallel path they are seen as, up to a certain distance, having a lower impedance when compared to the 'Cully' bus directly above. The gradient in correlation along the upper rows suggests this to be the case.

%Furthermore, in order to rule out the possibility that these patterns of de-correlation were caused by invalid data we can directly compare correlations developed over smaller window sizes. If we compare Figure \ref{weak6} to Figures \ref{datadrop} and \ref{misread}, we see that there are no PMUs that are entirely de-correlated or blacked-out (due to null data) -- both of which are distinct patterns of data events. We can thus safely classify this contingency as a power system contingency rather than a data issue.

\subsection{Bitmap Queries}
Queries were ran over the database to demonstrate the performance gains from analyzing and creating a bitmap index over the data. For these experiments, 4 million rows from the database were queried. The bitmap generated was compressed using a popular compression scheme known as Word-Aligned Hybrid (WAH)~\cite{Wu2004}. File Map was used to retrieve the records from the database once a query has been serviced. The bitmap results are compared against the common linear scan that is performed when searching a database as a basic comparison, and against MySQL, a popular database query engine that can store data collected from the PDC. Linear scanning has to scan every tuple in the database, therefore we know that the number of tuple it returns are the appropriate number of tuples that should be returned. We used this result to confirm the returned values from both Bitmap Indexing and MySQL.

\begin{table}[ht]
\small
\begin{center}
  \begin{tabular}{|c||p{4.8cm}|p{2.3cm}|p{2.3cm}|p{2.3cm}|p{2cm}|}  \hline
  \textbf{ID} & \textbf{Selection Criteria} & \textbf{Linear Scan (sec)}& \textbf{MySQL (sec)} &\textbf{Bitmap (sec)} &\textbf{Records Retrieved}\\
        \hline
        \hline
        1 & Find all records where PMU1 has a magnitude Voltage Magnitude of 533.
        & 25.859666 & 22.469 & 0.379387 & 160
        \\\hline

        2 & Find all records that occurred on exactly June 24, 2013 at 21:05 hours.
        & 25.350993 & 0.353 & 0.854952 & 3600
        \\\hline
        
        3 & Find all records that occurred on exactly June 24, 2013 at 21:06 hours.
        & 28.001001 & 0.396 & 0.922941 & 3600
        \\\hline
        
        4 & Find all records that occurred on exactly June 24, 2013 at 21:07 hours.
        & 26.133607 & 0.225 & 0.785588 & 3600
        \\\hline
        
        5 & Find all records that occurred on exactly June 24, 2013 at 21:06 hours with PMU having a Voltage Magnitude of 533.
        & 28.019449 & 0.046 & 0.001772 & 0
        \\\hline
        
        6 & Find all records in 2012.
        & 26.720291 & 23.714 & 0.0000601 & 0
        \\\hline
    \end{tabular}
\end{center}
\caption{\label{table:querycomp}Query Performance}
\end{table}

Table~\ref{table:querycomp} shows results from six queries that were run. When comparing MySQL and bitmap speeds, one should consider the language that was used as it will impact the performance. MySQL is implemented in \textit{C} while our bitmap indexing was implemented in \textit{Java}, which is in general slower. The SQL queries were performed with caching disabled, since we are interested in measuring the exact query execution time and not simply the data-fetch time. Query ID 1 is an example of a query where the user wishes to find when a specific PMU had a voltage magnitude of 533. An example of when this might happen is when the Correlation Visualization indicates that there is an event occurring when that PMU has a voltage magnitude of 533. The exact same query to the bitmap engine provides a $68\times$ and $60\times$ speed up on retrieval for linear scanning and MySQL respectively. Query IDs 2 through 4 demonstrate examples of requests for records at specific dates. These demonstrate that performing multiple queries with small adjustments does not require much additional time. Query IDs 5 and 6 shows queries for records that do not exist in the data set. Since the bitmap engine was able to examine the bit vector results without ever going to disk to see if the desired records are in the database, the speedup is many orders of magnitude greater than that of linear scanning. The bitmap query ID 5 takes slightly more time than ID 6 because ID 5 has to perform bitwise ANDs between each column, while ID 6 is simply checking a single column. There is very little time difference between the linear scan in ID 5 and 6.

MySQL outperforms bitmap indexing for a couple of reasons when searching for a specific date. For our database we used the \texttt{DATETIME} data column type in MySQL. This data type has a back end that builds a B$^{+}$-Tree~\cite{Comer:1979} index over it for efficient query processing. We can see whenever a query is submitted to search for a specific value of a PMU, even with a date specified, that bitmap indexing outperforms MySQL.

The linear scan times are so similar because no matter the query given, it is necessary to scan the entire data set to ensure accuracy. Bitmap index query times can vary and primarily depend on how many columns need to be compared and how many records need to be pulled from disk. In fact the majority of the time spent for the bitmap index queries is simply retrieving the records from disk, making I/O the limiting factor.

\section{Future Work}
The methods provided in this paper prove to be effective at data retrieval and show promising results with what events can be detected. Below are some directions we plan to take the implemented methods described in this paper.

Given large data sets, it is necessary to add additional methods of indexing for faster navigation and for queries to be returned in reasonable amounts of time.  One such method that could be applied is sampling. This adds tiers of bitmaps, i.e.,  bitmap indices for progressively more precise bitmaps, each one at a lower resolution of the data. For small amounts of data this is simply wasted space and too much overhead. When bit data such as this is introduced the sampling overhead begins to diminish as access times to the data doesn't scale up with the amount of data as quickly. 

%Bitmap indexing currently performs best when applied to a static database. To give the additional performance of allowing the bitmap to compress in real time can be advantageous. Increasing the frequency of compressing the new data would allow for queries resulting in recent tuples. Bitmap indices' high compression rates rely on large bit vectors. The tradeoff between compressing new data frequently and having more recent data is the compression ratio will be worse, leading to decreased query times. Understanding how often compression of new data and when re-compression of the entire database should take place will allow for the bitmap index to be updated effectively in real time.

Regarding the event detection algorithm, different clustering of PMU sites included in the correlation are being analyzed in order to optimize observability of system events in the visualization structure. Preliminary work indicates that when a subset of sites are correlated (\emph{e.g.} five PMUs sites - three ``close'' and two ``far'' relative to the system event), the correlation visuals clearly indicate sensitivity of the overall system correlation when analyzing phase angle. We will formalize this analysis with respect to other subsets of signals such as voltage magnitude, frequency, and rate-of-change of frequency. 

Window length plays a significant role in the detectability of events.  As noted in section \ref{sec:DataInputAndCorrelation}, short window lengths are well-suited for detecting data errors while longer window lengths are needed for detecting power system events.  It is desirable to use the shortest window length possible for detecting particular types of events in order to minimize computational power and time while also maintaining minimal false detection rates. As such, this relationship warrants further investigation. 

The event detection visualization can be scaled in multiple ways. For instance, if the PMU network is large enough then clustering the PMUs based on electrical distance and aggregating their outputs might be appropriate. To adjust how the visualization works currently on large scale PMU systems a ``scaled view'' could be implemented such that a portion of the entire matrix graphic could be enhanced, zooming the operator in on a particular area increasing visibility.

\label{sec:future}

\section{Conclusion}
\label{sec:conclusion}
We have shown that our system minimizes data driven bottlenecks that are typically associated with large-scale data sets. Specifically, compression of the index minimizes the space overhead, allowing it to be operated on within memory. Query response times are also minimized due to the utilization of indexing coupled with the FileMap structure. This results in the ability to perform frequent queries leading to efficient analysis of the data.

Additionally, our event detection algorithm demonstrates its capability in correlating PMU measurement readings resulting in effective monitoring of grid activity. This algorithm is coupled with a visualization component enabling grid operators the ability to efficiently identify occurrences of power system contingencies in addition to determining its location. This algorithm shows significant promise in transitioning to automated grid control. This level of intelligent computing is inevitable with the ever-increasing complexity of power generation, distribution, and consumption. We look forward to further developing this technology to advance the way that power engineers operate, control, and maintain the electric power grid.

%\todo[inline,color=green]{RB: Future work, we should be using negative sequence data.  It'll give stronger results when a system goes out of balance due to unsym faults, stronger than what we see with pos sequ}

\section*{Acknowledgment}
This research was funded in part by an Oregon BEST grant to E.~Cotilla-Sanchez, D.~Chiu, and R.~Bass. We thank the Bonneville Power Administration for providing the PMU data and support.

%\appendix

%\section{Section in Appendix}
%\label{appendix-sec1}

%% References


\bibliographystyle{elsarticle-num}

\bibliography{main}

\end{document}